\documentclass[aps,pra,twocolumn,
	amsmath,amssymb,amsfonts,floatfix,
	longbibliography,superscriptaddress,nobibnotes]{revtex4-2}

\usepackage[utf8]{inputenc}
\usepackage[T1]{fontenc} 
\usepackage{graphicx} 
\usepackage{bm}
\usepackage{placeins}

\usepackage[hypertexnames=false,colorlinks=true,citecolor=blue,
			linkcolor=blue,urlcolor=blue]{hyperref}

\usepackage{xfrac}
\usepackage[dvipsnames]{xcolor}
\definecolor{mygreen}{rgb}{0.0, 0.6, 0.0}
\definecolor{pjorange}{rgb}{0.8, 0.3, 0.0}
\definecolor{jlblue}{rgb}{0.2, 0.5, 0.7}

\graphicspath{{fig/}{./fig/}{.}}

\usepackage[normalem]{ulem}

\newcommand{\RNum}[1]{\uppercase\expandafter{\romannumeral #1\relax}}

\begin{document}

\title{Experimental electronic structure of the mineral superconductor covellite CuS}
   
\author{A.~Antezak}

\affiliation{Universit\'e Paris-Saclay, CNRS,  Institut des Sciences Mol\'eculaires d'Orsay, 
			91405, Orsay, France}
\author{T.~Kato}
\affiliation{Universit\'e Paris-Saclay, CNRS,  Institut des Sciences Mol\'eculaires d'Orsay, 
			91405, Orsay, France}
    \affiliation{Advanced Institute for Materials Research (WPI-AIMR), Tohoku University, Sendai 980-8577, Japan}

\author{P.~Rezende-Gonc\c alves}
\affiliation{Universit\'e Paris-Saclay, CNRS,  Institut des Sciences Mol\'eculaires d'Orsay, 
			91405, Orsay, France}
				
\author{F.~Fortuna}
\affiliation{Universit\'e Paris-Saclay, CNRS,  Institut des Sciences Mol\'eculaires d'Orsay, 
			91405, Orsay, France} 

\author{M.~Rosmus}

    \affiliation{Universit\'e Paris-Saclay, CNRS,  Institut des Sciences Mol\'eculaires d'Orsay, 
			91405, Orsay, France}
    \affiliation{National Synchrotron Radiation Centre SOLARIS, Jagiellonian University, Czerwone Maki 98, 30-392 Krak\'{o}w, Poland}
\author{N.~Olszowska}
    \affiliation{National Synchrotron Radiation Centre SOLARIS, Jagiellonian University, Czerwone Maki 98, 30-392 Krak\'{o}w, Poland}
			
\author{A.~F.~Santander-Syro}
\email{andres.santander-syro@universite-paris-saclay.fr}
\affiliation{Universit\'e Paris-Saclay, CNRS,  Institut des Sciences Mol\'eculaires d'Orsay, 
			91405, Orsay, France}

\author{E.~Frantzeskakis}
\email{emmanouil.frantzeskakis@universite-paris-saclay.fr}
\affiliation{Universit\'e Paris-Saclay, CNRS,  Institut des Sciences Mol\'eculaires d'Orsay, 
			91405, Orsay, France}

\date{\today}

\begin{abstract}
Covellite (CuS) is the first known natural mineral superconductor. Despite its simple chemical formula, covellite exhibits a rich crystal structure at the origin of several remarkable properties. The ionic arrangement in CuS crystals leads to a mixed valence of Cu and a second-order structural transition at 55 K. Despite the abundance of structural studies and theoretical reports on its electronic structure, there are scarce references on its experimental band structure. By means of Angle Resolved PhotoEmission Spectroscopy (ARPES), we have probed the experimental electronic structure of covellite. We compare our results with the predictions of density-functional theory (DFT) calculations. Our experimental data are in remarkable agreement with the calculations, revealing subtle fingerprints of the structural phase transition, and confirming the quasi-2D nature of the electronic structure of CuS.
\end{abstract}
%
\maketitle

\section{Introduction}

Covellite (CuS) is a copper sulfide mineral of notable scientific interest due to its potential applications in lithium battery cathodes \cite{CHUNG2002226, Jiang_cathode_2019}, visible-light photocatalysis \cite{basu_evolution_2010, kuchmii_catalysis_2001}, and plasmonics \cite{Yixin, Tiaoxing, Yi}. Most notably, it has earned the status of the first naturally occurring superconductor, with a critical temperature of 1.6 K \cite{di_benedetto2006, meissner_messungen_1929, GONCALVES20082742, Casaca_2012}.

Despite its simple chemical formula, CuS exhibits a structurally complex lattice featuring two inequivalent Cu and S sites, namely Cu(1), Cu(2) and S(1), S(2). These inequivalent sites form distinct Cu(2)-S$_4$ tetrahedra and Cu(1)-S$_3$ triangles [Fig. 1 (a)]. This complexity underpins an intermediate Cu valence state—averaging 1.33—with Cu(1) and Cu(2) atoms adopting valences of 1.5 and 1, respectively \cite{Kumar_2013, PEARCE20064635, Gainov_2009, Mazin_2012, morales-garcia_first-principles_2014}. This mixed valence bears resemblance to the physics of high-$T_c$ cuprate superconductors \cite{Mazumdar_2018, Proust_2019}, though high-temperature superconductivity in CuS has been ruled out in the absence of substantial hole doping \cite{RAVEAU2012291, Mazin_2012}. Interestingly, CuS undergoes a second-order transition below 55 K, during which the structural symmetry changes from hexagonal to orthorhombic through subtle atomic displacements and changes in Cu-S-Cu bonding angles \cite{Fjellvag_1988}.

Understanding the electronic structure of CuS is essential to provide new insights into the intermediate valence and its structural transition. Namely, Density Functional Theory (DFT) calculations have revealed the Fermi surface of CuS and the detailed band diagram that corresponds to the most widely accepted structure and to its intermediate valence \cite{Mazin_2012}. Moreover, other DFT studies, use the calculated electronic structure to propose microscopic mechanisms for the structural transition with theoretical models invoking van der Waals interactions, covalent S–S bonding, and Cu–S orbital mixing \cite{ LIANG1993405, morales-garcia_first-principles_2014, conejeros_nature_2014}. The aforementioned theoretical works suggested that the CuS conductivity is anisotropic due to the different overlap of in-plane and out-of-plane orbitals \cite{LIANG1993405, Mazin_2012, morales-garcia_first-principles_2014}, and that the band structure remains largely unchanged through the structural transition \cite{Mazin_2012, conejeros_nature_2014}

\begin{figure*}[t]
\includegraphics[width=1\linewidth]{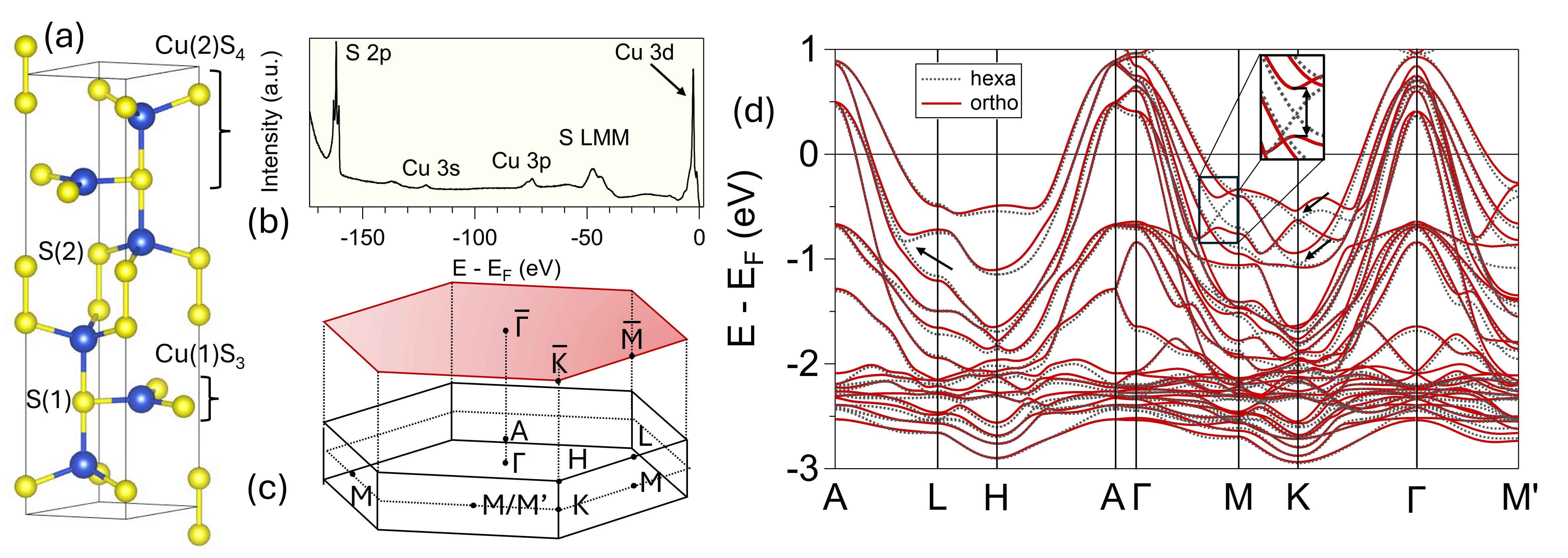}%
\caption{(a) Schematic model of the hexagonal high-temperature crystal structure of CuS. The crystal is composed of two inequivalent copper atoms (Cu(1) and Cu(2)) and two inequivalent sulfur atoms (S(1) and S(2)), represented respectively by blue and yellow spheres. (b) Angle-integrated photoemission spectrum measured with 200 eV photon energy and linear horizontal polarized light at 14 K. (c) High-temperature Brillouin zone (BZ) of CuS and its high-symmetry points. The red plane represents the projection of the BZ on the (001) plane, with the associated $\overline{\Gamma}, \overline{K}$ and $\overline{M}$ high-symmetry points. (d) Theoretical band structure of CuS calculated by DFT along a high-symmetry path, for the low-temperature orthorhombic (red) and high-temperature hexagonal (gray) structures. Because the low- and high-temperature BZ are almost identical, we use the same high-symmetry points in the hexagonal and orthorhombic structures. The inset shows in more detail the gap opening in the band structure of the orthorhombic phase along $\Gamma - M$, $\sim 750$ meV below $E_F$. Black arrows point at the different gap openings induced by the structural transition in the orthorhombic phase.}
\label{Fig1}
\end{figure*}

Despite the abundance of theoretical studies, experimental insights into the electronic structure of CuS remain limited. The only available data stem from early investigations —over three decades ago— based on Hall effect measurements \cite{NOZAKI1991306} and X-ray photoemission spectroscopy \cite{FOLMER1980153}. This highlights a striking imbalance between the numerous band structure calculations and the lack of direct experimental validation. In this context, probing the experimental electronic structure of CuS and comparing it with theoretical predictions is crucial to answer the questions addressed above: Is there a measurable impact of the structural transition on the electronic properties of CuS? Can the dimensionality of the electronic structure explain the anisotropic conductivity? Is the current theoretical framework experimentally justified? 

In this work, we address these questions by investigating the experimental electronic structure of CuS using Angle-Resolved Photoemission Spectroscopy (ARPES). The experimental results show remarkable agreement with our DFT calculations. Together, our experimental and theoretical findings uncover subtle, yet clear, spectroscopic fingerprints of the structural phase transition of CuS and shed new light into the dimensionality of its electronic structure.

\section{Methods}

\textbf{surface preparation:}  We used commercially-available CuS single crystals (2Dsemiconductors). To generate pristine surfaces, the single crystals were cleaved $\textit{in-situ}$ just before the ARPES experiments, exposing the (001) crystalline plane, and kept in UHV under a pressure better than 1$\times$10$^{-10}$~mbar. The cleavage surfaces exhibited multiple misaligned domains with respect to each other, with the size of the domains being comparable to the beam spot (60$\mu$m).\\

\textbf{ARPES experiments:} The ARPES measurements included in this article were carried out using a hemispherical electron analyzer with vertical slits at the URANOS beamline, SOLARIS Synchrotron (Poland). Typical experimental resolutions in energy and emission angle of photoelectrons were respectively 15~meV and 0.2$^{\circ}$. The electronic structure of the hexagonal and orthorhombic phases, was respectively measured at 65 K and 14 K. The temperature of each measurements is systematically indicated in the Figures caption. Unless stated otherwise, ARPES spectra shown in this manuscript were acquired in single-domain sample regions or in regions where the relative misalignment of neighboring domains was minimal.\\

\textbf{Density functional theory calculations:} The DFT calculations were performed under the Generalized Gradient Approximation (GGA) of the exchange-correlation energy, using the Quantum ESPRESSO integrated suite of open-source computer codes \cite{QE-2009, QE-2017, Giannozzi_2020}. We employed the Perdew, Burke, and Ernzerhof (PBE) functional for the exchange-correlation energy \cite{Perdew_1996}, along with the Projector Augmented Wave (PAW) pseudopotentials \cite{Blochl_1994, Kresse_1999} for Cu and S \cite{DALCORSO2014337}. The Monkhorst and Pack method was used to approximate the Brillouin zone integrals \cite{monkhorst_special_1976}. The energies were converged with respect to the $k$-point density (6 $\times$ 6 $\times$ 3), and a plane-wave cutoff energy of 1088 eV (80 Ry) was used to ensure convergence of the total energy within 10$^{-4}$ eV.
\begin{table}[h!]
\centering
\begin{tabular}{||c c c c c||} 
 \hline
  & Hexa exp. & Hexa rel. & Ortho exp. & Ortho rel. \\ [0.4ex] 
 \hline\hline
 a (\AA)& 3.782 & 3.804 & 3.763 & 3.813 \\ 
 b (\AA)& 3.782 & 3.804 & 6.586 & 6.574 \\
 c (\AA)& 16.289 & 16.511 & 16.225 & 16.443 \\
 V (\AA$^{3}$)& 204.13 & 206.958 & 201.053 & 206.108 \\
 E (meV) & $\emptyset$ & 0 & $\emptyset$ & $-$15 \\
 S(2)-S(2) & 2.03 & 2.13 & 2.04 & 2.12 \\ 
 Cu(2)-S(1) & 2.33 & 2.36 & 2.33 & 2.37 \\[1ex] 
 \hline
\end{tabular}
\caption{Comparative table of the two crystal structures with their experimental (exp.) and relaxed (rel.) parameters. The experimental values are those reported by Fjellv\aa{}g $\textit{et al}$ \cite{Fjellvag_1988}. The total energy per unit cell is obtained after the self-consistency field calculation and is referenced to the total energy of the relaxed hexagonal structure. The distances are given in \AA.}
\label{table_relaxation}
\end{table}

The experimental hexagonal and orthorhombic structures were relaxed to minimize the forces and stress in the unit cell. Table \ref{table_relaxation} compares the different parameters of the unit cell before and after relaxation. Our values after relaxation (2nd and 4th columns) are in good agreement with experimental results obtained by X-ray diffraction in the past literature \cite{Fjellvag_1988} (1st and 3rd columns), and are also in line with previous theoretical studies \cite{Mazin_2012}. After relaxation and self-consistent field calculations, the band structure plots were generated by computing the electronic structure along a high-symmetry path. For constant energy surface rendering, the electronic structure was calculated on a denser $k$-grid (24 $\times$ 24 $\times$ 6) and visualized using the FermiSurfer software \cite{KAWAMURA2019197}.\\

\textbf{2D-curvature details:} 2D-curvature rendering is used to enhance the intensity of weak or broad spectral features on the ARPES energy-momentum maps \cite{zhang_precise_2011}. For Figs. \ref{Fig3}(f) and \ref{Fig3}(g), boxcar smoothing was used eight times with a width of 0.004 \AA$^{-1}$ along the $k$-axis and four times with a width of 0.02 eV along the energy axis. The 2D-curvature factor a$_0$ was set equal to 1. In the case of Figs. \ref{Fig4}(c) and \ref{Fig4}(d), the data were initially smoothed using a Gaussian smoothing with full width at half maximum (FWHM) of 0.2 \AA$^{-1}$ and 0.055 eV and the 2D-curvature factor was set equal to 1$\times$10$^{-6}$. For Figs. \ref{Fig4}(e) and \ref{Fig4}(f), the FWHMs were 0.15 \AA$^{-1}$ and 0.04 eV with a 2D-curvature factor of 1$\times$10$^{-5}$.
Only negative values of the curvature are displayed. \\

\textbf{3D $k$-space mapping:} Within the free-electron final state model, ARPES measurements at constant photon energy $h\nu$ give the electronic structure at the surface of a spherical cap of radius $k = \sqrt{\frac{2m_{e}}{\hbar^{2} } }(h\nu - \Phi + V_{0})^{1/2}$. Here, $m_{e}$ is the free electron mass, $\Phi$ is the work function, and $V_{0}$ = 7 $\pm$ 1 eV is the inner potential of CuS, which was determined by the periodicity of the out-of-plane periodicity of the electronic structure (Fig. \ref{Fig4}). Measurements around normal emission provide the electronic structure in a plane nearly parallel to the surface plane. Likewise, measurements as a function of photon energy provide the electronic structure in a plane perpendicular to the surface.

\section{Results}

Fig. \ref{Fig1}(b) shows a wide-energy-range photoemission spectrum of CuS measured in the low-temperature phase at $T = 14$ K. All expected core-level photoemission peaks of Cu and S within the energy range of the spectrum are well resolved, indicating the purity of our sample. Moreover, one can readily observe a broader peak corresponding to the LMM Auger transition of sulfur \cite{COGHLAN1973317}. We note that no changes are expected in the photoemission spectrum of the core-level peaks at the high-temperature phase because the local chemical environment does not change significantly. There are only very minor changes in the bond lengths/angles and lattice constants \cite{Fjellvag_1988}. The most striking feature in Fig. \ref{Fig1}(b) is the line-shape of the S-$2p$ core level peak which is composed several individual peaks instead of the expected $2p_{1/2}$ and $2p_{3/2}$ doublet (see Fig. \ref{FigSup1} in Appendix). This is the signature of multiple chemical environments for the S atoms. As already mentioned, the crystal structure of CuS is made of triangular and tetrahedral Cu-coordinated blocks, forming respectively the so-called Cu(1)-S(1) and Cu(2)-S(2) planes, where the indices denote the different types of Cu and S atoms \cite{Fjellvag_1988}. The mere co-existence of such CuS$_4$ tetrahedra and CuS$_3$ triangular groups means that S atoms are indeed found in two different chemical environments in the bulk (i.e. S(1) and S(2) atoms). On top of two different bulk sites, symmetry breaking at the surface also creates a different chemical environment for the S atoms and may be responsible for the unidentified peaks in the lineshape of the S-$2p$ core level peak (see the Appendix for more details).

\begin{figure*}[t]
\centering
\includegraphics[scale=0.95]{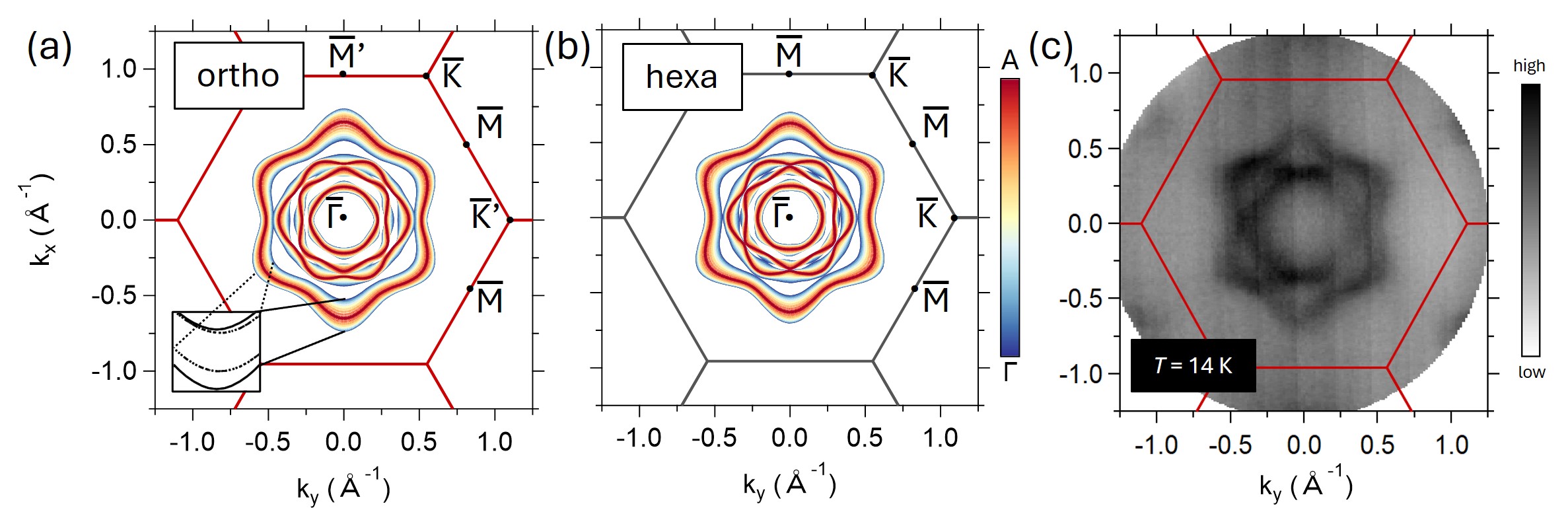}
\caption{(a), (b). Projection of the orthorhombic and hexagonal theoretical Fermi surface (FS) on the (001) plane. In the case of the orthorhombic FS (a), the sixfold symmetry is broken resulting in non-equivalent $\overline{M}$ and $\overline{M'}$ points, as well as non-equivalent $\overline{K}$ and $\overline{K'}$ points. The symmetry breaking at the $\overline{M'}$ and $\overline{M}$ points is shown in the inset, where the outermost contours pair at the $\overline{M'}$ (solid lines) and $\overline{M}$ (dashed lines) are depicted. The two contours are separated by a $k$-splitting of 0.13 \AA$^{-1}$ at the $\overline{M}$ point and 0.2 \AA$^{-1}$ at the $\overline{M'}$ point. The colorbar indicates the position along the [001] direction in the BZ of the projected contour. (c) Constant energy ($E - E_F = 0$) cut along the (001) plane, measured with 100 eV photon energy at 14 K in the orthorhombic phase. The cut was obtained by summing the data measured with linear vertical and horizontal polarized light.}
\label{Fig2}
\end{figure*}

We will now discuss the band structure of CuS. Fig. \ref{Fig1}(c) shows the hexagonal Brillouin zone (BZ) of the high-temperature phase. Due to the very small lattice distortion and the minor changes in the unit cell across the transition, the BZ in the low-temperature orthorhombic phase is almost identical to the high-temperature hexagonal BZ (Fig. \ref{FigSup2} for more details). Thus, henceforth, we will use the same high-symmetry points to describe the band structure of CuS in the hexagonal and orthorhombic phases as has been routinely done in past studies \cite{Mazin_2012}. Note, however, that, in the low-temperature orthorhombic phase, not all the $\overline{M}$ points are equivalent, and this will play an important role in later discussions. We will hereafter let $\overline{M}'$ denote the high-symmetry points in the direction where the sixfold symmetry is lifted.

\begin{figure*}[t]
\centering
\includegraphics[scale=0.5]{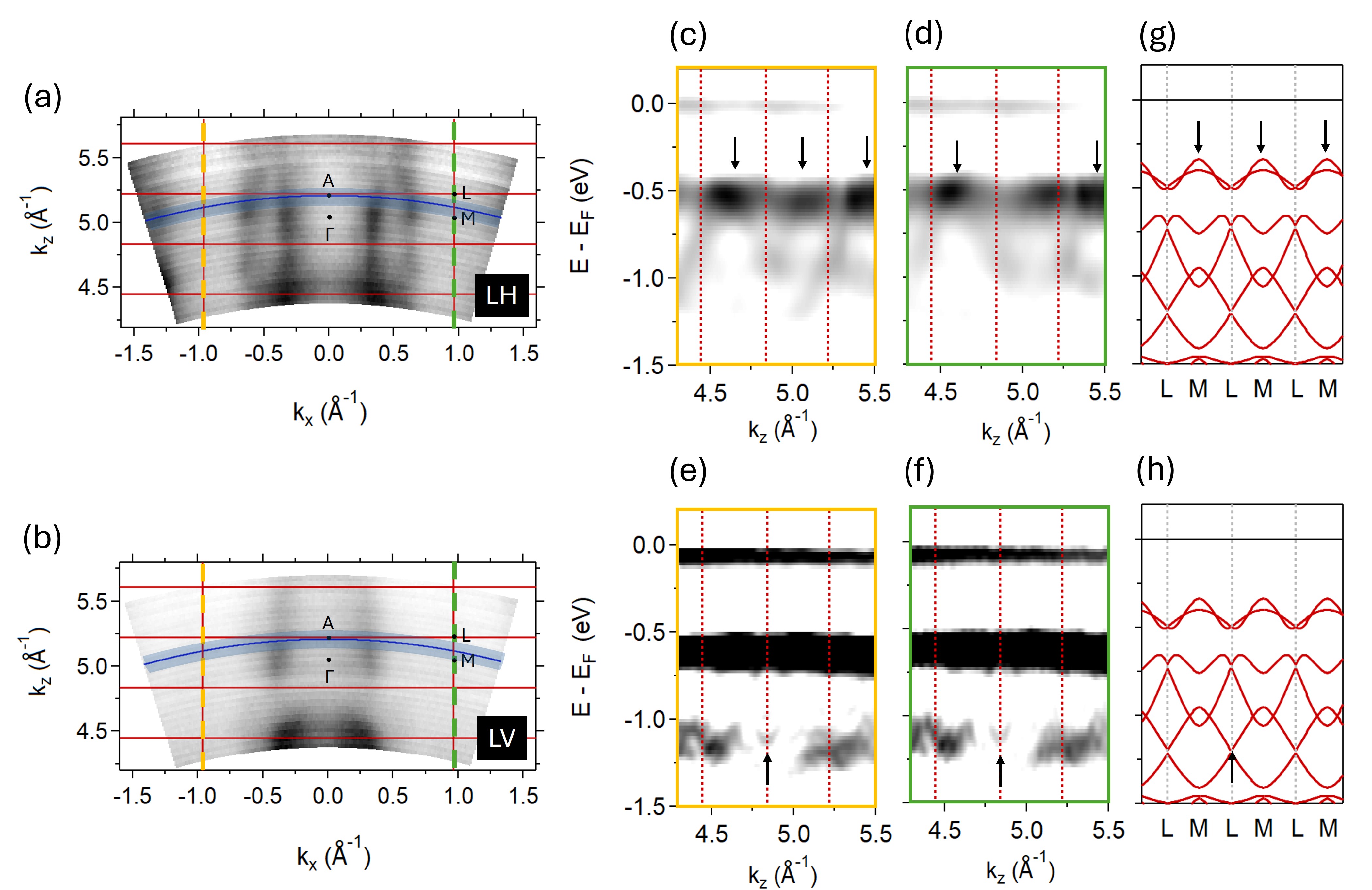}
\caption{(a), (b) Constant energy ($E - E_F = 0$) cut, measured at 14 K by systematically changing the photon energy from 70 to 120 eV in steps of 1 eV in order to probe the out-of-plane electronic structure, using linear horizontal (a) and vertical (b) polarized light. The superimposed red lines denote the borders of several BZs spanning the reciprocal space. The blue arc represents the 100 eV photon energy line, energy used for the in-plane measurements. The width of this arc gives the uncertainty $\delta k_z = 1/\lambda$, where $\lambda$ is the photoelectron escape depth \cite{strocov_intrinsic_2003}. The yellow and green dashed lines indicate the position of the cuts displayed in (c), (d) and (e), (f). (c), (e) 2D-curvature of the energy-momentum dispersion extracted by cutting respectively in (a) and (b) along the negative edge of the BZ, i.e along the $M - L$ direction. (d), (f) Same as (c) and (e) but the cuts are made along the positive edge. (g) Theoretical band structure along $M - L$. (h) Same as (g). Arrows mark some experimental features that can be easily associated to the DFT bands.}
\label{Fig4}
\end{figure*}

Fig. \ref{Fig1}(d) presents the calculated band structure of CuS in its high-temperature hexagonal phase (gray) and low-temperature orthorhombic phase (red), along the high-symmetry directions of the hexagonal BZ. In our calculations, the small atomic rearrangements across the transition are detailed in Table \ref{table_relaxation}, while the calculated cell expansion of 0.5\% is slightly smaller than the experimentally verified expansion \cite{Fjellvag_1988}. The electronic structure remains very similar across the structural transition with only few differences which are manifested as avoided band crossings in the low-temperature phase. Such energy gap openings in the low-temperature phase are observed at $E - E_F \sim$ $-500$ meV along $\Gamma - M$ and at $K$, as well as at around $-750$ meV along $A - L$, all indicated by the arrows. The gap openings in the low-temperature band structure of CuS might be related to the reduced symmetry of the crystal structure in the orthorhombic phase. The breaking of sixfold symmetry in the orthorhombic phase also results in the electronic structure along $\Gamma - M$ being different to the electronic structure along $\Gamma - M'$. According to our calculations, along $\Gamma - M$ at $E - E_F \sim$ $-500$ meV, there is an energy gap of approximately $300$ meV, while along $\Gamma - M'$, the energy gap is absent. There are therefore tiny differences in the band structure of the orthorhombic phase depending whether one is probing the $\Gamma - M$ or the $\Gamma - M'$ high-symmetry direction. We further note that the band structure along $\Gamma - M$ predicts a relatively flat state at $E - E_F \sim$ $-750$ meV, as shown in the inset. This state defines the lower border of the band-gap and will be discussed later in more detail, in the context of Fig. 4. Lastly, one can note the relatively similar band structure in the $A - L - H$ (BZ border) and in the $\Gamma - M - K$ (BZ center) planes, indicating the absence of significant energy-momentum dispersion along the out-of-plane direction of the BZ.
Our theoretical calculations are in agreement with previous studies \cite{Mazin_2012, conejeros_nature_2014, LIANG1993405}, which also show that the structural transition has a minor effect on the electronic structure.

\begin{figure*}[t]
\centering
\includegraphics[scale=0.65]{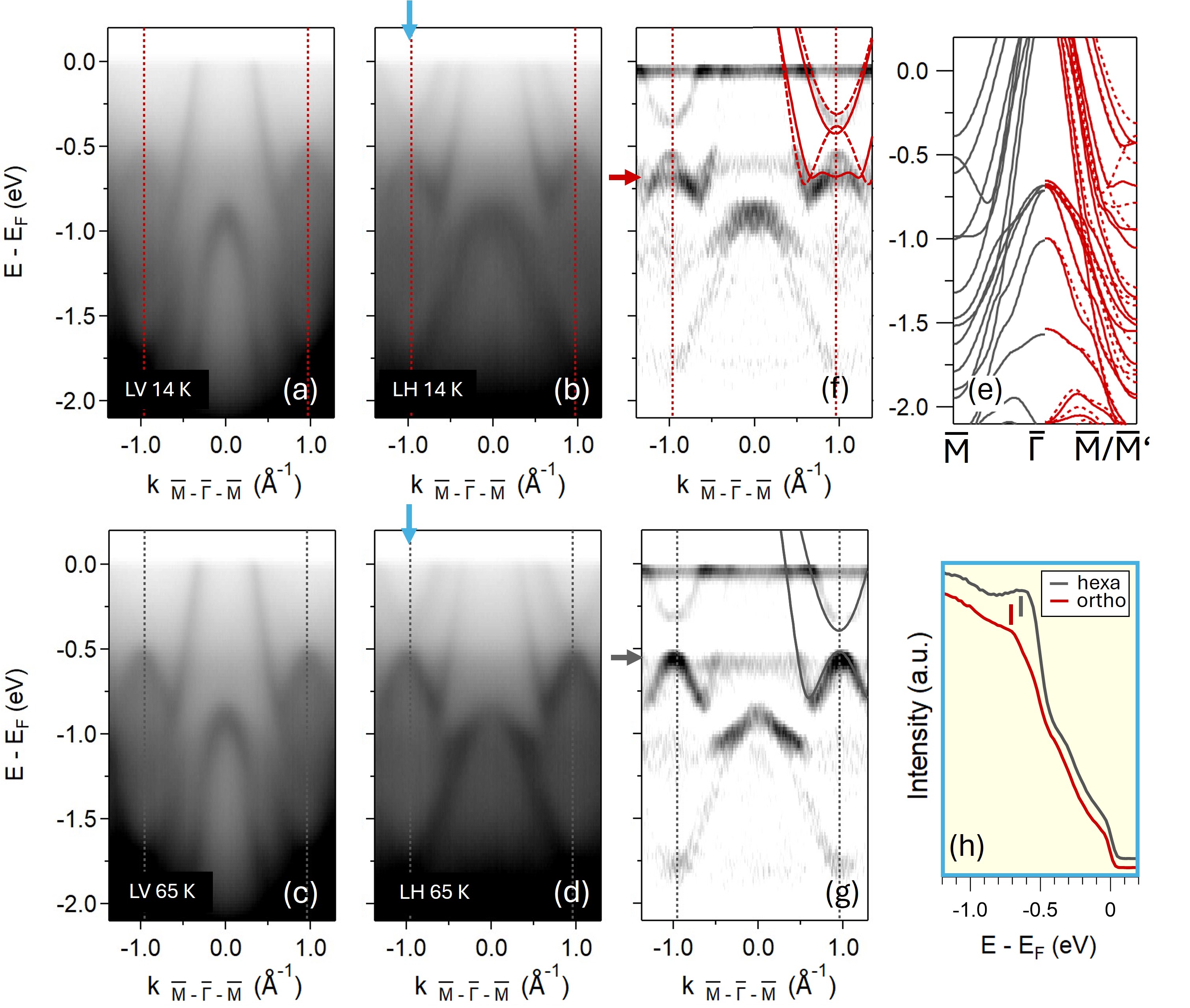}
\caption{(a), (b) Energy-momentum dispersion maps measured along $\overline{M} - \overline{\Gamma} - \overline{M}$ with 100 eV photon energy and using linear vertical (a) and horizontal (b) polarized light, at 14 K in the orthorhombic phase. (c), (d) Same as (a) and (b) but measured at 65 K in the hexagonal phase. The blue arrows in (b) and (d) indicate the position of the Energy Distribution Curves (EDCs) displayed in (h). (f), (g) Data in (b) and (d) treated with the 2D-curvature method. The red and gray arrows pinpoint the position of the peaks identified in the EDC. On top of the curvature data, selected calculated bands along $\overline{\Gamma} - \overline{M'}$ and $\overline{\Gamma} - \overline{M}$ are superimposed. In panel (f) the data was first symmetrized along the $k = 0$ line. The dashed lines are selected calculated bands along $\overline{\Gamma} - \overline{M}$ which are also superimposed on the experimental data. (e) Calculated band structure along the $\overline{\Gamma} - \overline{M}$ direction, with an out-of-plane momentum value of 0.09675 \AA$^{-1}$, i.e. in the middle of $\Gamma - A$ and corresponding to $k_z$ momentum measured with a photon energy of 100 eV, for the orthorhombic (right / red) and hexagonal (left / gray) phases. On the right part, dashed lines denote the calculated band structure along $\overline{\Gamma} - \overline{M}'$ at the same $k_z$ value in the orthorhombic phase. (h) EDCs measured along the zone boundary in (b) and (d). The EDCs reveal two different peak positions for the two different phases.}
\label{Fig3}
\end{figure*}

The theoretical band structure shown in Fig. \ref{Fig1}(d) reveals that there is a total of eight bands crossing the Fermi level around $\Gamma$, which should then give rise to eight distinct Fermi contours centered at $\Gamma$. Fig. \ref{Fig2}(a) displays the (001)-surface-projected Fermi surface of CuS in the low-temperature orthorhombic phase. Each of the four projected theoretical contours corresponds to a pair of two closely-separated states, which are degenerate at $A$, as seen in Fig. \ref{Fig1}(d), but clearly separated ---although still very closely spaced--- at $\Gamma$. Therefore, the slight dependence of the in-plane Fermi surface contours on the different values of the $k_z$ out-of-plane momentum, results in a apparent broadening of all these in-plane Fermi surfaces upon their projection onto the surface Brillouin zone. The color scale indicates the out-of-plane momentum value ---ranging from the zone center $\Gamma$ to the zone boundary $A$--- at which the in-plane Fermi surface contour has been extracted from the 3D electronic structure. A careful inspection of the projected contour reveals that the $k_z$-broadening in the orthorhombic phase is not isotropic. Due to the twofold symmetry of the orthorhombic structure, the momentum splitting along the $k_x$ direction [i.e. oriented vertically in Fig. \ref{Fig2}(a)] is larger than in the other equivalent directions [see inset in Fig. \ref{Fig2}(a)]. This results in enhanced broadening along these directions, thus breaking the sixfold symmetry of the contours and resulting in a slightly elongated flower-shaped feature along the vertical direction. Symmetry breaking will be more clearly shown with the energy-momentum dispersions discussed in the context of Fig. \ref{Fig3}. The anisotropic distribution of the contour broadening is not evident for the other pairs of contours. Finally, Fig. \ref{Fig2}(b) shows the equivalent of Fig. \ref{Fig2}(a) but now in the high-temperature hexagonal phase. No major changes in the Fermi contours are observed in the high-temperature phase [Fig. \ref{Fig2}(b)] with respect to its low-temperature counterpart, except for the fact that contour broadening in the hexagonal phase strictly follows the sixfold symmetry.


We now turn to the experimental electronic structure of CuS. Figure \ref{Fig2}(c) shows the experimental (001)-Fermi surface of CuS measured at 14 K, thus in the low-temperature, orthorhombic phase. The Fermi surface is composed of closed contours of sixfold symmetry, centered around $\overline{\Gamma}$. The outer contours have a flower-like shape, while the inner are hexagonal. A comparison with the projected Fermi surface obtained by DFT calculations and shown in Fig. \ref{Fig2}(a) reveals four sixfold contours whose overall shape is in very good agreement with the experimental results. We attribute the two outer contours to the ones observed experimentally. We note that previous theoretical investigations revealed that the orbital character of the two outer contours stems from the Cu(2)-S(2) layer, while the two inner contours are predominantly related to the Cu(1)-S(1) layer \cite{conejeros_nature_2014}. Given that the two inner contours were not observed under any experimental conditions that we tested (variable photon energy from 60 to 200 eV and variable photon polarization), we cannot attribute their absence to the matrix elements of the photoemission process \cite{moser2017}. We therefore invoke the possibility that our ARPES experiments might be preferentially sensitive to the Cu(2)-S(2) layer.

An interesting aspect of the electronic structure of CuS is its dimensionality. Already, while discussing the theoretical Fermi surface contours in Figs. \ref{Fig2}(a) and \ref{Fig2}(b), we noted that the $k_z$-dependence of the electronic structure results in a slight broadening of the surface-projected contours, but it does not modify the bands in an important way. This is due to the absence of significant energy-momentum dispersion along the out-of-plane direction of the BZ, as predicted by the DFT calculations shown in Fig. \ref{Fig1}(d) and in previous studies \cite{morales-garcia_first-principles_2014, Mazin_2012, LIANG1993405}. We experimentally established the dimensionality of the CuS electronic structure by means of photon-energy-dependent ARPES. Figs. \ref{Fig4}(a) and \ref{Fig4}(b) show the out-of-plane constant energy ($E - E_{F}$ = 0) cuts measured, respectively, with linear horizontal and vertical polarized light. Unlike the in-plane Fermi surface map composed of several closed contours (Fig. \ref{Fig2}(c)), the out-of-plane maps of Figs. \ref{Fig4}(a) and \ref{Fig4}(b) are fully open quasi-parallel sheets, thereby signifying the absence of an out-of-plane dispersion near the Fermi level, indicative of 2D electronic behavior.

Although the electronic structure around the Fermi level is clearly two-dimensional, this conclusion cannot be simply extended to all energies. For instance, Fig. \ref{Fig1}(d) predicts dispersing bands along the out-of-plane $\Gamma - A$ direction in the energy range between $-$1 and $-$2 eV. The poor out-of-plane momentum resolution of ARPES \cite{strocov_intrinsic_2003} makes it challenging to experimentally measure a small $k_z$ dispersion in systems like CuS, which have a large unit cell along $z$ and hence a small out-of-plane constant of the BZ. Nevertheless, we did observe clear signs of an out-of-plane dispersion at lower energies. Figs. \ref{Fig4}(c)-(f) show the experimental energy-momentum dispersions along $M - L$. Several features, marked by the arrows, can be linked to our DFT calculations shown in Figs. \ref{Fig4}(g) and \ref{Fig4}(h). 
Through the observed out-of-plane periodicity of the electronic structure and its comparison to DFT predictions, we could estimate the inner potential of CuS with an approximate value of 7 $\pm$ 1 eV. This implies that a photon energy of 100 eV (used in Figs. \ref{Fig2} and \ref{Fig4}) yields photoelectrons with a $k_z$ corresponding to the high-symmetry point $A$ in reciprocal space (i.e. to the BZ boundary). In conclusion, the electronic structure of CuS is multi-dimensional: while the Fermi surface consists of states with two-dimensional character, electronic states located deeper in energy exhibit a distinct three-dimensional nature. The two-dimensionality of the Fermi surface, as revealed by our ARPES measurements, is consistent with the anisotropic conductivity of CuS reported in previous theoretical studies \cite{LIANG1993405, Mazin_2012, morales-garcia_first-principles_2014}\\

In order to thoroughly describe the electronic structure of CuS across the transition, we will now focus on the experimental band structure along the $\overline{\Gamma} - \overline{M}$ high-symmetry direction. Figures \ref{Fig3}(a) and \ref{Fig3}(b) show the energy-momentum dispersion at low temperature (orthorhombic phase) measured with photons of linear vertical (LV) and linear horizontal (LH) polarization, respectively. Figures \ref{Fig3}(c) and \ref{Fig3}(d) show the equivalent energy-momentum dispersion at 65 K, in the hexagonal phase. As can be seen in Figs.  \ref{Fig3}(a) and \ref{Fig3}(c) there are two hole-like states at $\overline{\Gamma}$: one is crossing the Fermi level, and the other exhibits a maximum at an approximate binding energy of 1 eV. The presence of metallic hole bands is in agreement with early magneto-transport results that had predicted metallic hole conduction in CuS \cite{NOZAKI1991306}. At the zone boundary ($\overline{M}$), there is another concave parabolic state with a band maximum just below $-$500 meV. Figs. \ref{Fig3}(b) and \ref{Fig3}(d) show complementary data to the above: two parabolic electron-like states are centered at $\overline{M}$ and cross the Fermi level. Those states were not probed with LV light. Moreover, the hole-like state with a maximum at $-$1 eV has a different effective mass in Figs. \ref{Fig3}(a) and \ref{Fig3}(c), compared to the one observed in Figs. \ref{Fig3}(b) and \ref{Fig3}(d). Therefore, these two states are of different origin and correspond to different bands in the multiplet of the closely spaced theoretical bands shown in Fig. \ref{Fig3}(e).

A quick inspection of the theoretical band structure in Fig. \ref{Fig3}(e), confirms that although all experimental bands can be linked to their theoretical counterparts, not every theoretical band can be experimentally probed. As noted previously, the metallic hole-like states that give rise to the two innermost contours around $\overline{\Gamma}$ [Figs. \ref{Fig2}(a) and \ref{Fig2}(b)] cannot be observed in our ARPES experiment. In order to enhance the features of the experimental electronic structure shown in Figs. \ref{Fig3}(b) and \ref{Fig3}(d), we have plotted the 2D-curvature of the original photoemission data in Figs. \ref{Fig3}{(f) and \ref{Fig3}(g). Additionally, the data shown in Fig. \ref{Fig3}(f) has been symmetrized around the center of the BZ. The curvature data sheds further light on the experimental band structure. The detailed energy-momentum dispersion of the highest-lying experimental band at $\Gamma$ reveals that it is rather a doublet as seen in Figs. \ref{Fig3}(f) and \ref{Fig3}(g).
We now focus on the electronic structure around $\overline{M}$. In the orthorhombic phase [Fig. \ref{Fig3}(f)], there is a relative flat state with a band maximum at around $-$700 meV at $\overline{M}$, as indicated by the red arrow. This flat state has no counterpart in the hexagonal phase [Fig. \ref{Fig3}(g)], where no state is experimentally observed below the hole-like band. These observations are highlighted in Fig. \ref{Fig3}(h), which displays the Energy Distribution Curves (EDCs) measured at the $\overline{M}$ point in the two structural phases. We observe a clear energy shift, $\Delta E \sim80$ meV, of the peak position towards lower energies in the orthorhombic phase. These experimentally observed differences call for further analysis and a careful comparison of the theoretical band structure in the two phases.

On the data measured below (above) the hexagonal-to-orthorhombic structural transition we have superimposed the experimentally observed DFT bands calculated in the orthorhombic (hexagonal) phase. The results are, respectively, shown in Figs. \ref{Fig3}(f) and \ref{Fig3}(g). In the high-temperature phase [Fig. \ref{Fig3}(g)] the agreement of theory and experiment is satisfactory even though DFT slightly underestimates the bottom of the highest-lying band. This small discrepancy is not surprising since DFT calculations within the GGA PBE framework can systematically underestimate the size of band gaps \cite{jonesDensityFunctionalTheory2015, grumetQuasiparticleApproximationFully2018, leeFirstprinciplesApproachPseudohybrid2020}. However, the interpretation of the low-temperature data [Fig. \ref{Fig3}(f)] is less straightforward. On one hand, DFT can efficiently reproduce the quasi-flat band that was discussed in the previous paragraph and has no counterpart in the hexagonal phase. On the other hand, there is an experimental hole-like band at $\overline{M}$ that has no counterpart in the DFT bands along $\overline{\Gamma} - \overline{M}$ of the orthorhombic phase. We remind the reader that the phase transition breaks the sixfold symmetry, which yields inequivalent $\overline{M}$ and $\overline{M'}$ points. In Fig. \ref{Fig3}(f), we also superimpose in dashed lines the calculated DFT bands along $\overline{\Gamma} - \overline{M'}$. The experimental band, which did not have theoretical counterpart, is now captured by DFT, although it overestimates its band top (see explanation before and \cite{jonesDensityFunctionalTheory2015, grumetQuasiparticleApproximationFully2018, leeFirstprinciplesApproachPseudohybrid2020}). We explain the presence of the signatures of the two inequivalent directions in our energy-momentum maps by the presence of misaligned orthorhombic domains probed under the beam spot. We observed the tendency of CuS single crystals to form misaligned domains when put under stress (upon cleaving). Moreover, although we are able to probe single domains in the hexagonal phase after cleavage, the structural transition induces additional stress, resulting in three different potential orientations for the orthorhombic domains.
Since the beam spot size is of the order of $\sim$ 60$\mu$m, it is most probable that it probes multiple misaligned orthorhombic domains during the ARPES experiments. Given the very minor differences in the Fermi surface of CuS around the inequivalent $\overline{M}$ and $\overline{M'}$ points, it is not surprising that we could not observe the symmetry breaking in the experimental Fermi surface [Fig. \ref{Fig2}(c)], but only in high-resolution energy-momentum dispersion at lower energies.

Despite the complications to fully explain the low-temperature experimental data of CuS, we note that ---to the best of our knowledge--- there had been so far no other experimental report on differences in its electronic structure above and below the transition. We therefore hope that our work will motivate future studies on CuS that will further discuss the evolution of its electronic structure across its structural phase transition. 

\section{Conclusions}

Our ARPES measurements show that the experimental electronic structure of CuS single crystals is in very good agreement with our DFT calculations. Consistent with theoretical predictions, the electronic structure is quasi-two-dimensional, featuring open Fermi surface contours along the out-of-plane direction. In contrast, the in-plane Fermi surface consists of multiple hole-like pockets that remain largely unaffected by the structural phase transition. In the low-temperature orthorhombic phase, we identify spectroscopic signatures of the transition, including the emergence of a band at the $\overline{M}$ high-symmetry point that has no counterpart in the high-temperature hexagonal phase.
This study aims to address the long-standing lack of experimental data on the electronic structure of CuS and serve as a foundation for future experimental research on covellite.

\acknowledgments 
Work at ISMO was supported by public grants from the French National Research Agency (ANR), projects SUPERNICKEL No. ANR-21-CE30-0041-05 and MICROVAN No. ANR-24-CE30-3213-01, and by the CNRS International Research Project EXCELSIOR. T.K. acknowledges support from GP-Spin at Tohoku University and JSPS. This publication was partially developed under the provision of the Polish Ministry and Higher Education project "Support for research and development with the use of research infra-structure of the National Synchrotron Radiation Centre SOLARIS” under contract no 1/SOL/2021/2.

\section{Appendix}

\renewcommand{\thefigure}{S\arabic{figure}}
\setcounter{figure}{0}
\section*{S-$2p$ core level peak}
\begin{figure}[h]
\centering
\includegraphics[scale=0.4]{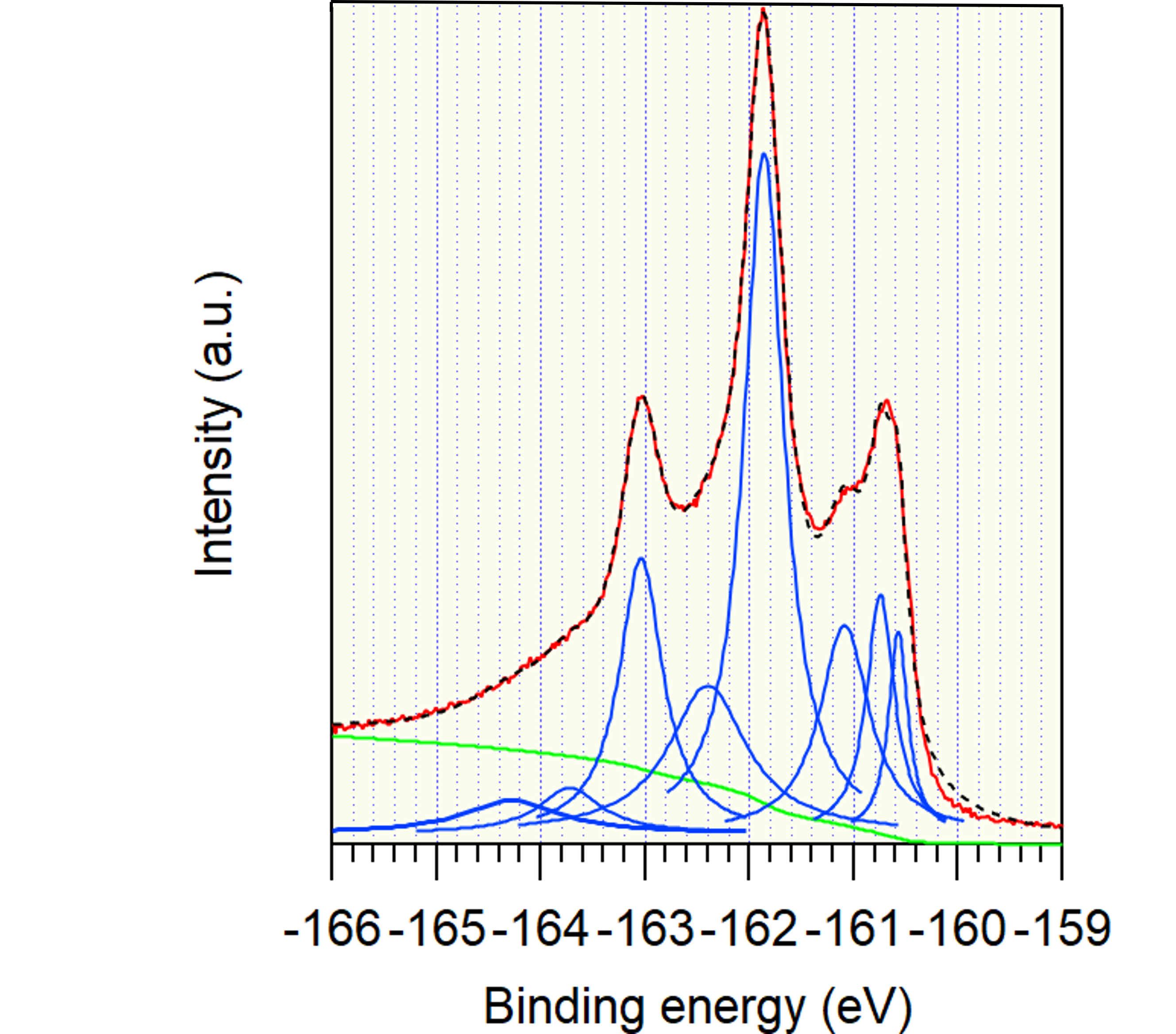}
\caption{Angle-integrated photoemission spectrum of the S-$2p$ peak, measured with 200 eV photon energy and linear horizontal polarized light. The raw signal is plotted in red, while the black dashed line corresponds to the sum of the Shirley background (green) and the eight lorentzian peaks (blue).}
\label{FigSup1}
\end{figure}
Fig. \ref{FigSup1} displays the line-shape of the S-$2p$ peak measured by means of angle-integrated photoemission with 200 eV photon energy. The main doublet, composed of the peaks at $-$161.9 eV and $-$163 eV, corresponds to the tabulated S-$2p$ doublet, namely S-$2p_{3/2}$ and S-$2p_{1/2}$, respectively, and originates from one of the two inequivalent atomic sites in the bulk [i.e. S(1) and S(2) atoms]. We identified a second doublet composed of the peaks at $-$162.4 eV and $-$163.7 eV \cite{thompson_x-ray_2001}. Although their energy splitting (1.3 eV) slightly deviates from the expected value of 1.1 eV, we could attribute this doublet to the second inequivalent S site in the bulk. The set of three peaks et around $-$160.5 eV might be related to surface S-S bonds \cite{Liu_2024}. Lastly, the broad and weakly intense peak at $-$164.3 eV remains unidentified and might be related to other surface defects.

\section*{Reciprocal space}
\begin{figure}[h]
\centering
\includegraphics[scale=0.6]{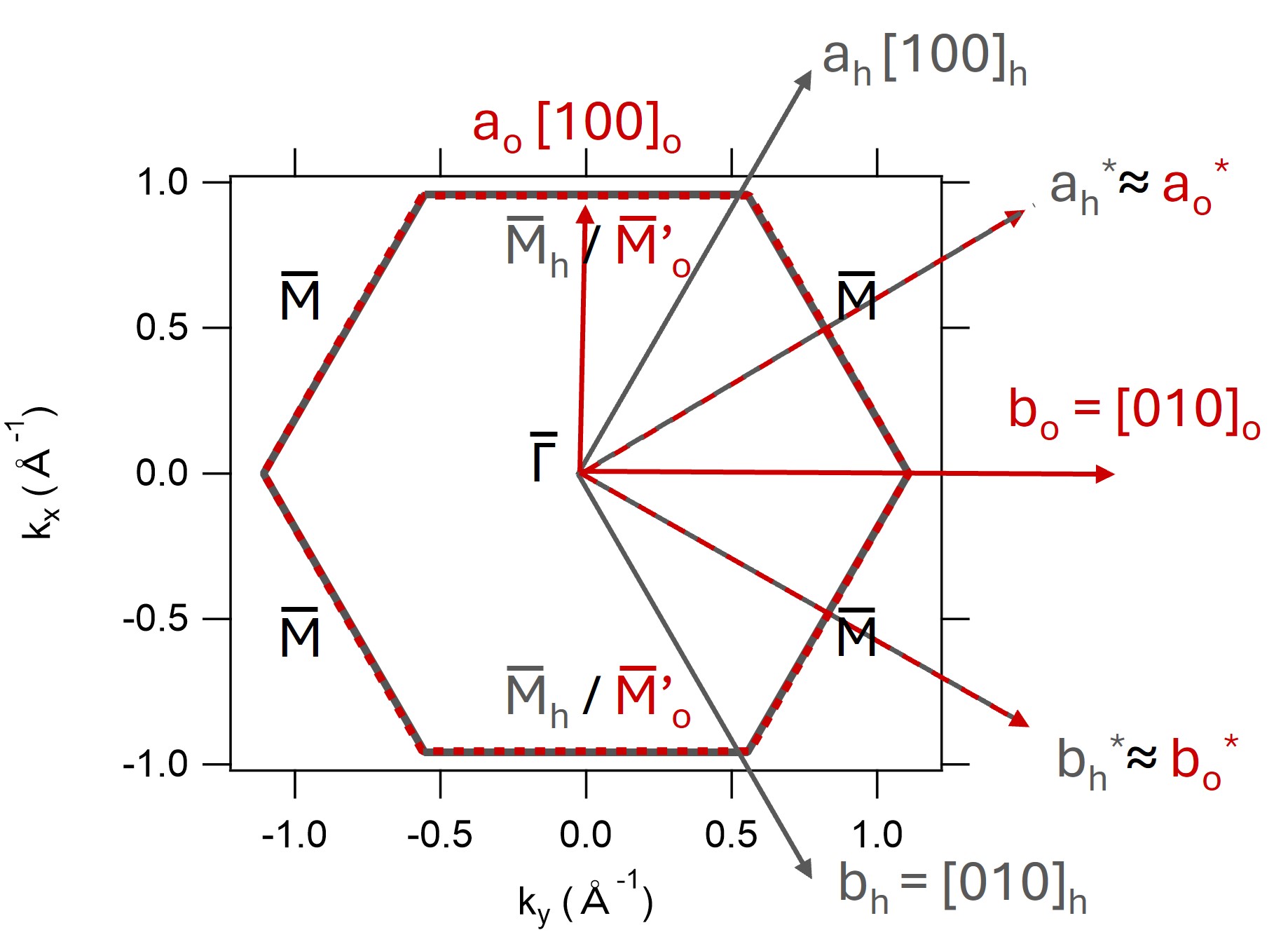}
\caption{The surface projected Brillouin zones of CuS in the hexagonal (gray) and orthorhombic (red) phases. High-symmetry points, as well as real-space and reciprocal-space unit vectors are included in the figure and explained in the text.}
\label{FigSup2}
\end{figure}
Fig. \ref{FigSup2} presents, respectively, the high- (gray) and low-temperature (red) (001)-surface projected BZ of CuS. The projected Brillouin zones are calculated with Igor programming language macros using the lattice parameters of the relaxed structures. The two BZs are almost identical although derived from two different real-space vector bases, namely $(a_o ,b_o)$ for the orthorhombic and $(a_h ,b_h)$ for the hexagonal BZ. These sets of vectors are used to derive the reciprocal-space unit vectors $(a_o^* ,b_o^*)$ and $(a_h^* ,b_h^*)$, which are almost identical. Therefore, besides the $\overline{M'_o}$ point being different from the $\overline{M_h}$ point for symmetry reasons, one can keep the same labeling to describe the BZ of CuS in both structural phases.

\setcounter{enumiv}{0}


\begin{thebibliography}{10}

\bibitem{CHUNG2002226}
J.-S. Chung and H.-J. Sohn.
\newblock Electrochemical behaviors of {CuS} as a cathode material for lithium
  secondary batteries.
\newblock {\em Journal of Power Sources}, 108(1):226--231, 2002.

\bibitem{Jiang_cathode_2019}
Kyle Jiang, Zonghai Chen, and Xiangbo Meng.
\newblock {CuS} and {Cu$_{2}$S} as cathode materials for lithium batteries: A
  review.
\newblock {\em ChemElectroChem}, 6(11):2825--2840, 2019.

\bibitem{basu_evolution_2010}
Mrinmoyee Basu, Arun~Kumar Sinha, Mukul Pradhan, Sougata Sarkar, Yuichi
  Negishi, Govind ~, and Tarasankar Pal.
\newblock Evolution of {Hierarchical} {Hexagonal} {Stacked} {Plates} of {CuS}
  from {Liquid}-{Liquid} {Interface} and its {Photocatalytic} {Application} for
  {Oxidative} {Degradation} of {Different} {Dyes} under {Indoor} {Lighting}.
\newblock {\em Environmental Science \& Technology}, 44(16):6313--6318, August
  2010.
\newblock Publisher: American Chemical Society.

\bibitem{kuchmii_catalysis_2001}
S.~Ya. Kuchmii, A.~V. Korzhak, A.~E. Raevskaya, and A.~I. Kryukov.
\newblock Catalysis of the {Sodium} {Sulfide} {Reduction} of {Methylviologene}
  by {CuS} {Nanoparticles}.
\newblock {\em Theoretical and Experimental Chemistry}, 37(1):36--41, January
  2001.

\bibitem{Yixin}
Yixin Zhao, Hongcheng Pan, Yongbing Lou, Xiaofeng Qiu, JunJie Zhu, and Clemens
  Burda.
\newblock Plasmonic {Cu$_{2-x}$S} nanocrystals: Optical and structural
  properties of copper-deficient copper({\RNum{1}}) sulfides.
\newblock {\em Journal of the American Chemical Society}, 131(12):4253--4261,
  2009.
\newblock PMID: 19267472.

\bibitem{Tiaoxing}
Tiaoxing Wei, Yufeng Liu, Wenjing Dong, Yun Zhang, Chanyan Huang, Yan Sun, Xin
  Chen, and Ning Dai.
\newblock Surface-dependent localized surface plasmon resonances in {CuS}
  nanodisks.
\newblock {\em ACS Applied Materials \& Interfaces}, 5(21):10473--10477, 2013.
\newblock PMID: 24138006.

\bibitem{Yi}
Yi~Xie, Luigi Carbone, Concetta Nobile, Vincenzo Grillo, Stefania D?Agostino,
  Fabio Della~Sala, Cinzia Giannini, Davide Altamura, Christian Oelsner, Carola
  Kryschi, and P.~Davide Cozzoli.
\newblock Metallic-like stoichiometric copper sulfide nanocrystals: Phase- and
  shape-selective synthesis, near-infrared surface plasmon resonance
  properties, and their modeling.
\newblock {\em ACS Nano}, 7(8):7352--7369, 2013.
\newblock PMID: 23859591.

\bibitem{di_benedetto2006}
Francesco Di~Benedetto, Miria Borgheresi, Andrea Caneschi, Guillaume Chastanet,
  Curzio Cipriani, Dante Gatteschi, Giovanni Pratesi, Maurizio Romanelli, and
  Roberta Sessoli.
\newblock First evidence of natural superconductivity: covellite.
\newblock {\em European Journal of Mineralogy}, 18(3):283--287, 07 2006.

\bibitem{meissner_messungen_1929}
W.~Meissner.
\newblock Messungen mit {Hilfe} von flussigem {Helium}. {V}.
  {Supraleitfahigkeit} von {Kupfersulfid}.
\newblock {\em Zeitschrift fur Physik}, 58(7):570--572, July 1929.

\bibitem{GONCALVES20082742}
A.P. Goncalves, E.B. Lopes, A.~Casaca, M.~Dias, and M.~Almeida.
\newblock Growth of {CuS} platelet single crystals by the high-temperature
  solution growth technique.
\newblock {\em Journal of Crystal Growth}, 310(11):2742--2745, 2008.

\bibitem{Casaca_2012}
A~Casaca, E~B Lopes, A~P Goncalves, and M~Almeida.
\newblock Electrical transport properties of {CuS} single crystals.
\newblock {\em Journal of Physics: Condensed Matter}, 24(1):015701, dec 2011.

\bibitem{Kumar_2013}
Prashant Kumar, Rajamani Nagarajan, and Ritimukta Sarangi.
\newblock Quantitative {X}-ray absorption and emission spectroscopies:
  electronic structure elucidation of {Cu$_{2}$S} and {CuS}.
\newblock {\em J. Mater. Chem. C}, 1:2448--2454, 2013.

\bibitem{PEARCE20064635}
C.I. Pearce, R.A.D. Pattrick, D.J. Vaughan, C.M.B. Henderson, and G.~{van der
  Laan}.
\newblock Copper oxidation state in chalcopyrite: Mixed {Cu} {$d^{9}$} and
  {$d^{10}$} characteristics.
\newblock {\em Geochimica et Cosmochimica Acta}, 70(18):4635--4642, 2006.

\bibitem{Gainov_2009}
R.~R. Gainov, A.~V. Dooglav, I.~N. Pen'kov, I.~R. Mukhamedshin, N.~N. Mozgova,
  I.~A. Evlampiev, and I.~A. Bryzgalov.
\newblock Phase transition and anomalous electronic behavior in the layered
  superconductor {CuS} probed by {NQR}.
\newblock {\em Phys. Rev. B}, 79:075115, Feb 2009.

\bibitem{Mazin_2012}
I.~I. Mazin.
\newblock Structural and electronic properties of the two-dimensional
  superconductor {CuS} with 1$\frac{1}{3}$-valent copper.
\newblock {\em Phys. Rev. B}, 85:115133, Mar 2012.

\bibitem{morales-garcia_first-principles_2014}
A.~Morales-Garcia, Antonio Lenito~Jr. Soares, Egon~C. Dos~Santos, Heitor~A.
  de~Abreu, and Helio~A. Duarte.
\newblock First-{Principles} {Calculations} and {Electron} {Density}
  {Topological} {Analysis} of {Covellite} ({CuS}).
\newblock {\em The Journal of Physical Chemistry A}, 118(31):5823--5831, August
  2014.
\newblock Publisher: American Chemical Society.

\bibitem{Mazumdar_2018}
Sumit Mazumdar.
\newblock Valence transition model of the pseudogap, charge order, and
  superconductivity in electron-doped and hole-doped copper oxides.
\newblock {\em Phys. Rev. B}, 98:205153, Nov 2018.

\bibitem{Proust_2019}
Cyril Proust and Louis Taillefer.
\newblock The remarkable underlying ground states of cuprate superconductors.
\newblock {\em Annual Review of Condensed Matter Physics}, 10(Volume 10,
  2019):409--429, 2019.

\bibitem{RAVEAU2012291}
B.~Raveau, Tapati Sarkar, and A.~Pautrat.
\newblock Comment on our article ?superconducting-like behaviour of the
  layered chalcogenides {CuS} and {CuSe} below 40 {K}? (solid state sciences
  13 (2011) 1874?1878).
\newblock {\em Solid State Sciences}, 14(2):291, 2012.

\bibitem{Fjellvag_1988}
Fjellv\aa{}g Helmer, Gr\o{}nvold Frederik, and St\o{}len Svein.
\newblock Low-temperature structural distortion in {CuS}.
\newblock {\em Zeitschrift fur Kristallographie}, 184(1-2):111--121, 1988.

\bibitem{LIANG1993405}
W.~Liang and M.-H. Whangbo.
\newblock Conductivity anisotropy and structural phase transition in covellite
  {CuS}.
\newblock {\em Solid State Communications}, 85(5):405--408, 1993.

\bibitem{conejeros_nature_2014}
Sergio Conejeros, Iberio de P.~R. Moreira, Pere Alemany, and Enric Canadell.
\newblock Nature of {Holes}, {Oxidation} {States}, and {Hypervalency} in
  {Covellite} ({CuS}).
\newblock {\em Inorganic Chemistry}, 53(23):12402--12406, December 2014.
\newblock Publisher: American Chemical Society.

\bibitem{NOZAKI1991306}
Hiroshi Nozaki, Kenji Shibata, and Naoki Ohhashi.
\newblock Metallic hole conduction in {CuS}.
\newblock {\em Journal of Solid State Chemistry}, 91(2):306--311, 1991.

\bibitem{FOLMER1980153}
J.C.W Folmer and F~Jellinek.
\newblock The valence of copper in sulphides and selenides: An {X-ray}
  photoelectron spectroscopy study.
\newblock {\em Journal of the Less Common Metals}, 76(1):153--162, 1980.

\bibitem{QE-2009}
Paolo Giannozzi, Stefano Baroni, Nicola Bonini, Matteo Calandra, Roberto Car,
  Carlo Cavazzoni, Davide Ceresoli, Guido~L Chiarotti, Matteo Cococcioni,
  Ismaila Dabo, Andrea {Dal Corso}, Stefano de~Gironcoli, Stefano Fabris, Guido
  Fratesi, Ralph Gebauer, Uwe Gerstmann, Christos Gougoussis, Anton Kokalj,
  Michele Lazzeri, Layla Martin-Samos, Nicola Marzari, Francesco Mauri,
  Riccardo Mazzarello, Stefano Paolini, Alfredo Pasquarello, Lorenzo Paulatto,
  Carlo Sbraccia, Sandro Scandolo, Gabriele Sclauzero, Ari~P Seitsonen,
  Alexander Smogunov, Paolo Umari, and Renata~M Wentzcovitch.
\newblock {QUANTUM} {ESPRESSO}: a modular and open-source software project for
  quantum simulations of materials.
\newblock {\em Journal of Physics: Condensed Matter}, 21(39):395502 (19pp),
  2009.

\bibitem{QE-2017}
P~Giannozzi, O~Andreussi, T~Brumme, O~Bunau, M~Buongiorno Nardelli, M~Calandra,
  R~Car, C~Cavazzoni, D~Ceresoli, M~Cococcioni, N~Colonna, I~Carnimeo, A~Dal
  Corso, S~de~Gironcoli, P~Delugas, R~A~DiStasio Jr, A~Ferretti, A~Floris,
  G~Fratesi, G~Fugallo, R~Gebauer, U~Gerstmann, F~Giustino, T~Gorni, J~Jia,
  M~Kawamura, H-Y Ko, A~Kokalj, E~Kucukbenli, M~Lazzeri, M~Marsili,
  N~Marzari, F~Mauri, N~L Nguyen, H-V Nguyen, A~Otero de-la Roza, L~Paulatto,
  S~Ponce, D~Rocca, R~Sabatini, B~Santra, M~Schlipf, A~P Seitsonen,
  A~Smogunov, I~Timrov, T~Thonhauser, P~Umari, N~Vast, X~Wu, and S~Baroni.
\newblock Advanced capabilities for materials modelling with {QUANTUM}
  {ESPRESSO}.
\newblock {\em Journal of Physics: Condensed Matter}, 29(46):465901, 2017.

\bibitem{Giannozzi_2020}
Paolo Giannozzi, Oscar Baseggio, Pietro Bonfa, Davide Brunato, Roberto Car,
  Ivan Carnimeo, Carlo Cavazzoni, Stefano de~Gironcoli, Pietro Delugas,
  Fabrizio Ferrari~Ruffino, Andrea Ferretti, Nicola Marzari, Iurii Timrov,
  Andrea Urru, and Stefano Baroni.
\newblock Quantum {ESPRESSO} toward the exascale.
\newblock {\em The Journal of Chemical Physics}, 152(15):154105, 2020.

\bibitem{Perdew_1996}
John~P. Perdew, Kieron Burke, and Matthias Ernzerhof.
\newblock Generalized gradient approximation made simple.
\newblock {\em Phys. Rev. Lett.}, 77:3865--3868, Oct 1996.

\bibitem{Blochl_1994}
P.~E. Bl\"ochl.
\newblock Projector augmented-wave method.
\newblock {\em Phys. Rev. B}, 50:17953--17979, Dec 1994.

\bibitem{Kresse_1999}
G.~Kresse and D.~Joubert.
\newblock From ultrasoft pseudopotentials to the projector augmented-wave
  method.
\newblock {\em Phys. Rev. B}, 59:1758--1775, Jan 1999.

\bibitem{DALCORSO2014337}
Andrea {Dal Corso}.
\newblock Pseudopotentials periodic table: From {H} to {Pu}.
\newblock {\em Computational Materials Science}, 95:337--350, 2014.

\bibitem{monkhorst_special_1976}
Hendrik~J. Monkhorst and James~D. Pack.
\newblock Special points for {Brillouin}-zone integrations.
\newblock {\em Physical Review B}, 13(12):5188--5192, 1976.

\bibitem{KAWAMURA2019197}
Mitsuaki Kawamura.
\newblock Fermisurfer: Fermi-surface viewer providing multiple representation
  schemes.
\newblock {\em Computer Physics Communications}, 239:197--203, 2019.

\bibitem{zhang_precise_2011}
P.~Zhang, P.~Richard, T.~Qian, Y.-M. Xu, X.~Dai, and H.~Ding.
\newblock A precise method for visualizing dispersive features in image plots.
\newblock {\em Review of Scientific Instruments}, 82(4):043712, April 2011.

\bibitem{COGHLAN1973317}
W.A. Coghlan and R.E. Clausing.
\newblock Auger catalog calculated transition energies listed by energy and
  element.
\newblock {\em Atomic Data and Nuclear Data Tables}, 5(4):317--469, 1973.

\bibitem{strocov_intrinsic_2003}
V.N. Strocov.
\newblock Intrinsic accuracy in 3-dimensional photoemission band mapping.
\newblock {\em Journal of Electron Spectroscopy and Related Phenomena},
  130(1-3):65--78, July 2003.

\bibitem{moser2017}
Simon Moser.
\newblock An experimentalist's guide to the matrix element in angle resolved
  photoemission.
\newblock {\em J. Electron Spectrosc. Relat. Phenom.}, 214:29--52, January
  2017.

\bibitem{jonesDensityFunctionalTheory2015}
R.~O. Jones.
\newblock Density functional theory: {{Its}} origins, rise to prominence, and
  future.
\newblock {\em Rev. Mod. Phys.}, 87(3):897--923, August 2015.

\bibitem{grumetQuasiparticleApproximationFully2018}
Manuel Grumet, Peitao Liu, Merzuk Kaltak, Ji{\v r}{\'i} Klime{\v s}, and Georg
  Kresse.
\newblock Beyond the quasiparticle approximation: {Fully} self-consistent
  {$GW$} calculations.
\newblock {\em Phys. Rev. B}, 98(15):155143, 2018.

\bibitem{leeFirstprinciplesApproachPseudohybrid2020}
Sang-Hoon Lee and Young-Woo Son.
\newblock First-principles approach with a pseudohybrid density functional for
  extended {{Hubbard}} interactions.
\newblock {\em Phys. Rev. Research}, 2(4):043410, December 2020.

\bibitem{thompson_x-ray_2001}
A.C. Thompson.
\newblock {\em {X-ray} {Data} {Booklet}}.
\newblock Lawrence Berkeley National Laboratory, University of California,
  2001.

\bibitem{Liu_2024}
Jianan Liu, Xuemeng Sun, Yuying Fan, Yaoguang Yu, Qi~Li, Jing Zhou, Huiquan Gu,
  Keying Shi, and Baojiang Jiang.
\newblock {P}-{N} {H}eterojunction {E}mbedded {C}u{S}/{T}i{O}2 {B}ifunctional
  {P}hotocatalyst for synchronous hydrogen production and benzylamine
  conversion.
\newblock {\em Small}, 20(10):2306344, 2024.

\end{thebibliography}
\end{document}